\documentclass[aps, superscriptaddress]{revtex4-1}
\pdfoutput=1
\usepackage{graphicx}

\usepackage{amsmath}
\usepackage{amssymb}
\usepackage{bm}
\usepackage{color}

\newcommand{\pa}{\partial}

\newcommand{\mean}[1]{\langle{#1}\rangle}

\newcommand{\bra}[1]{\langle{#1}|}
\newcommand{\ket}[1]{|{#1}\rangle}

\newcommand{\Tr}{{\rm Tr}\hspace{0.07cm}}

\newcommand{\abs}[1]{{|#1|}}

\newcommand{\argmax}{\mathop{\rm arg~max}\limits}

\definecolor{darkgreen}{rgb}{0,0.4,0} % R(赤),G(緑),B(青)

\usepackage{ulem}
\begin{document}
	\title{Quantum spin van der Pol oscillator - a spin-based limit-cycle oscillator exhibiting quantum synchronization}
	\author{Yuzuru Kato}
	\email{Corresponding_author: katoyuzu@fun.ac.jp}
	\affiliation{Department of Complex and Intelligent Systems, Future University Hakodate, Hokkaido 041-8655, Japan }	
	\author{Hiroya Nakao}
	\affiliation{Department of Systems and Control Engineering
	and Research Center for Autonomous Systems Materialogy,%, Institute of Innovative Research,
		Institute of Science Tokyo, Tokyo 152-8552, Japan}
	\date{\today}

\begin{abstract}
We introduce a quantum spin van der Pol (vdP) oscillator as a prototypical model of quantum spin-based limit-cycle oscillators, which coincides with the quantum optical vdP oscillator in the high-spin limit. The system is described as a noisy limit-cycle oscillator in the semiclassical regime at large spin numbers, exhibiting frequency entrainment to a periodic drive. Even in the smallest spin-1 case, mutual synchronization, Arnold tongues, and entanglement tongues in two dissipatively coupled oscillators, and collective synchronization in all-to-all coupled oscillators are clearly observed. The proposed quantum spin vdP oscillator will provide a useful platform for analyzing quantum spin synchronization.
\end{abstract}

\maketitle

%%%%%%%%%%%%%%%%%
%%% Section 1 %%%
%%%%%%%%%%%%%%%%%
%
\section{Introduction}
Rhythmic oscillations and synchronization are ubiquitous in real-world systems,
including laser oscillations, spiking neurons, flashing fireflies, and mechanical vibrations
~\cite{winfree2001geometry, kuramoto1984chemical, ermentrout2010mathematical, pikovsky2001synchronization, glass1988clocks, strogatz1994nonlinear}. With recent advances in quantum technology, quantum synchronization in spin-$1$ atoms \cite{laskar2020observation}, in nuclear spin systems \cite{krithika2022observation}, and on the IBM Q system \cite{koppenhofer2020quantum} have been 
experimentally realized, and numerous theoretical studies on quantum synchronization have been conducted
~\cite{
	lee2013quantum,
	walter2014quantum,
	sonar2018squeezing,
	lorch2016genuine,
	lee2014entanglement,
	walter2015quantum,
	es2020synchronization,  
	bastidas2015quantum,
	kato2019semiclassical,
	mok2020synchronization,
	roulet2018synchronizing,
	roulet2018quantum,
	koppenhofer2019optimal, 
	xu2014synchronization,
	xu2015conditional,
	kato2022definition, kato2023quantum, 
	solanki2023symmetries, solanki2022role, 
	chia2020relaxation, arosh2021quantum, 
	galve2017quantum, buvca2022algebraic, 
	schmolke2022noise, tan2022half, 
	eneriz2019degree, 
	jaseem2020generalized, 
	tindall2020quantum,
	cabot2019quantum,cabot2021metastable, nadolny2023macroscopic}.

The classical van der Pol (vdP) oscillator, 
originally proposed for analyzing nonlinear oscillations in triode circuits \cite{van1927vii},  stands as a paradigmatic model of limit-cycle oscillators.  
Recently, the quantum van der Pol oscillator \cite{lee2013quantum}, which we call the {\it quantum optical vdP oscillator} hereafter, was proposed as a prototypical model of quantum limit-cycle oscillators in optical systems; it describes a harmonic oscillator coupled with two types of dissipation, i.e., single-photon gain and two-photon loss, which are quantum analogs of the negative damping and nonlinear damping in the classical vdP oscillator, respectively.
Using the quantum optical vdP oscillator, novel features of quantum synchronization have been investigated theoretically,  
e.g., entrainment by a harmonic drive \cite{lee2013quantum, walter2014quantum} and by squeezing \cite{sonar2018squeezing},
multiple phase locking in the strong quantum regime \cite{lorch2016genuine}, mutual synchronization \cite{lee2014entanglement, walter2015quantum}, and effects of quantum measurements \cite{es2020synchronization}.

Since small spin systems are suitable for analyzing many-body quantum synchronization 
\cite{tindall2020quantum} and experimental realization~\cite{laskar2020observation, koppenhofer2020quantum}, quantum synchronization in spin-$1$ systems has also attracted considerable attention recently~\cite{roulet2018synchronizing, roulet2018quantum}. Roulet and Bruder highlighted the spin-$1$ (three-level) system as the smallest model suitable for formulating quantum synchronization \cite{roulet2018synchronizing}, while several studies have indicated that spin-$1/2$ (two-level) systems can also exhibit quantum synchronization \cite{zhirov2008synchronization, cabot2019quantum, parra2020synchronization}. Such discrepancy seems to arise
because (i) stable limit-cycle oscillations, (ii) frequency entrainment to a periodic external drive, (iii) mutual frequency synchronization between two oscillators, and (iv) collective synchronization in globally (all-to-all) coupled oscillators, which are essentially important
features in the analysis of classical synchronization  \cite{pikovsky2001synchronization},  have not been studied 
systematically for a single, specific model of quantum spin systems.

In quantum optical systems,  the quantum optical vdP oscillator plays a pivotal role in the theoretical analysis of quantum synchronization.
Limit-cycle oscillations under the effect of quantum noise in the semiclassical regime have been analyzed using the quasiprobability distribution in the phase-state representation \cite{lee2013quantum}. Frequency entrainment of the oscillator to a periodic external drive \cite{walter2014quantum} and mutual frequency synchronization between two oscillators \cite{walter2015quantum} have been observed through the alignment of the peak frequency of the oscillator's power spectrum to the periodic drive or to each other \cite{walter2014quantum, walter2015quantum}. Also, collective synchronization in a population of oscillators with global coupling has been studied using the mean-field analysis
\cite{lee2014entanglement}. 
	Such systematic analysis of the essential features of quantum synchronization is enabled by the quantum optical van der Pol (vdP) oscillator, which, in the classical limit, reproduces the Stuart-Landau oscillator (normal form of the supercritical Hopf bifurcation), the most fundamental and simplest model of a classical limit-cycle oscillator.
	Owing to this correspondence with the simplest normal form of a classical limit-cycle oscillator, we can directly compare synchronization behavior between quantum optical systems and classical systems.
For quantum spin systems, however, no such universal model as the quantum optical vdP oscillator exists.

In this study,  we propose a {\it quantum spin vdP oscillator} as a 
fundamental model of self-sustained oscillators in quantum spin systems.
We prove that it coincides with the quantum optical vdP {oscillator} in the high-spin limit and yields the Stuart-Landau oscillator. 
Through a systematic numerical analysis, we reveal that the quantum spin vdP oscillators exhibit all essential features of quantum synchronization, i.e., frequency entrainment, mutual synchronization, and collective synchronization,
and we compare the synchronization  
behavior among quantum spin systems, quantum optical systems, and classical systems.
In particular, we report the first explicit observation of mutual frequency synchronization between a pair of single quantum spin-based limit-cycle oscillators with the smallest spin-1 systems. 

%%%%%%%%%%%%%%%%%
%%% Section 2 %%%
%%%%%%%%%%%%%%%%%

%
\section{Spin coherent state}

In this section, we give an overview of the spin coherent state.
Following Arecchi {\it et al.}~\cite{arecchi1972atomic}, we consider a spin-$J$ system consisting of an assembly of $N$ spin-$1/2$ atoms under a constant magnetic field in the $z$ direction  and introduce the (collective) angular-momentum operators 
\begin{align}
	\label{eq:angular}
	&J_{\mu} =\frac{1}{2} \sum_{n=1}^{N} \sigma_{\mu}^{n} \quad(\mu=x, y, z),
	\quad
	J_{\pm}=\sum_{n=1}^{N} \sigma_{\pm}^{n},
	\quad
	{\bm J}^{2}=J_{x}^{2}+J_{y}^{2}+J_{z}^{2},
\end{align}
satisfying
\begin{align}
	&\bm{J}^{2}|J, m\rangle =J(J+1)|J, m\rangle,
	\quad
	J_{\pm}\left|J, m\right\rangle = \sqrt{J(J+1)-m\left(m \pm 1 \right)}\left|J, m \pm 1 \right\rangle,
	\cr
	&J_{z}|J, m\rangle =m|J, m\rangle,
	\quad
	J_x = (J_+  + J_-)/2, ~ J_y = -i(J_+  - J_-)/2,
\end{align}
where $\sigma_{\mu=x, y, z}^{n}$ are the Pauli matrices for the $n$th atom. This spin-$J$ system ($J = 1/2, 1, 3/2, \ldots$) is described by a $(2J + 1)$-dimensional subspace spanned by the degenerate eigenstates of ${\bm J}^2$ associated with the eigenvalue $J(J+1)$, i.e., 
$\ket{J,J}, \ket{J,J-1}, \cdots, \ket{J,-(J-1)},  \ket{J,-J}$.

\begin{figure} [!htbp]
	\begin{center}
		\includegraphics[width=0.7\hsize,clip]{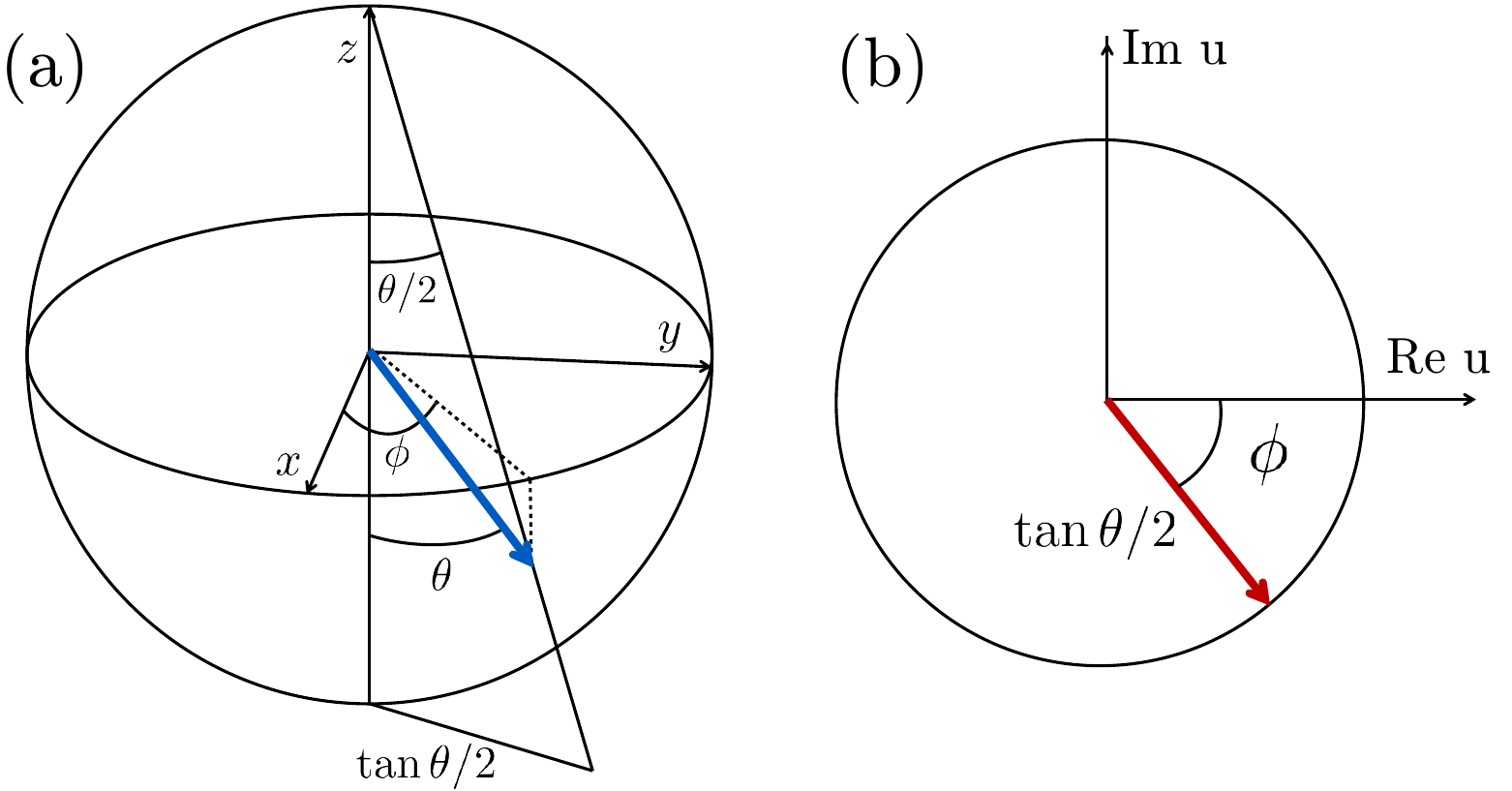}
		\caption{
			Mapping of the point $\ket{\theta, \phi}$ on the unit sphere characterizing a spin coherent state to the point $(u, u^*)$ on the complex plane.
		}
		\label{fig_1}
	\end{center}
\end{figure}

The spin coherent state \cite{radcliffe1971some, arecchi1972atomic}, which is an analog of the standard coherent state 
in quantum optical systems~\cite{gardiner1991quantum, carmichael2007statistical}, is defined as 
\begin{align}
	\label{eq:scs_sm}
	\ket{\theta, \phi} = R_{\theta, \phi}\ket{J, -J},
	\quad
	R_{\theta, \phi} = e^{-i\theta (J_{x} \sin \phi - J_{y} \cos \phi) },
\end{align}
where the spin coherent state is specified by a point $(\theta, \phi)$ ($0 \leq \theta < \pi$, $0 \leq \phi \leq 2\pi$)  on the unit sphere. This point $(\theta, \phi)$  can be mapped to a point $(u, u^*)$ on the complex plane as shown in Fig.~\ref{fig_1} by introducing $u = \tan(\theta/2) e^{-i\phi}$, and the spin coherent state can be rewritten as
\begin{align}
	\label{eq:scs_u_sm}
	\ket{u} =
	\ket{\theta, \phi} = 
	\sum_{m=-J}^{J} \binom{2J}{J+m}^{1/2} 
	\frac{u^{m+J}}{(1 + \abs{u}^2)^J} \ket{J, m}.
\end{align}
The spin coherent states satisfy the following completeness relation:
\begin{align}
	\frac{ ( 2 J + 1 ) }{ 4 \pi } \int d\theta d\phi\sin  \theta \ket{\theta,\phi} \bra{\theta,\phi}
	= \frac{ ( 2 J + 1 ) }{ \pi } \int \frac{d^2 u}{( 1 +  \abs{u}^2)^2} \ket{u}\bra{u} = \bm{1},
\end{align}
where $\bm{1}$ is the $(2J+1) \times (2J+1)$ identity matrix. 

We can also introduce the Hushimi $Q$ functions on the unit sphere \cite{gilmore1975classical} and on the complex plane as 
\begin{align}
	Q(\theta, \phi) = \frac{2J+1}{4\pi}\bra{\theta, \phi} \rho \ket{\theta, \phi} ,
	\quad
	Q(u, u^{*}) = \frac{2J+1}{\pi}\bra{u} \rho \ket{u},
\end{align}
respectively, which satisfy
\begin{align}
	\int d\theta d\phi \sin \theta Q(\theta, \phi) = 
	\int d^2 u \bar{Q}(u, u^{*}) = 1,
	\quad
	\bar{Q}(u, u^{*}) = \frac{Q(u, u^{*})}{( 1 +  \abs{u}^2)^2},
\end{align}
where we introduced $ \bar{Q}(u, u^{*})$ that includes the coefficient arising from the coordinate transformation.

\section{Quantum spin vdP oscillators}
Our main proposal in this study is the {\it quantum spin vdP oscillator} described by the following quantum master equation for the density matrix $\rho$:
\begin{align}
	\label{eq:qsvdp}
	\dot{\rho}
	&=
	-i [H, \rho] 
	+ \gamma_{1} \mathcal{D}[J'_{+}]\rho
	+ \gamma_{2} \mathcal{D}[(J'_{-})^2]\rho,
	\cr
	&
	H = \omega J_{z},
	\quad
	\mathcal{D}[L]\rho = L \rho L^{\dag} - 
	\frac{1}{2}\left(\rho L^{\dag} L + L^{\dag} L \rho \right),
\end{align}
where $H$ is the Hamiltonian, $\omega$ gives the natural frequency of the oscillator determined by the spin magnetic moment and the intensity of the magnetic field, $\gamma_{1}$ and $\gamma_{2}$ are the intensities of the negative damping and nonlinear damping, respectively, and $\mathcal{D}[L]$ represents the Lindblad form. Here, we have introduced the rescaled operators $J'_{+} = J_{+}/\sqrt{2J}$ and $J'_{-} = J_{-}/\sqrt{2J}$. Note that we need to assume $J \geq 1$, because the nonlinear damping term in Eq.~(\ref{eq:qsvdp}) vanishes when $J = 1/2$ 
since $(J_{-})^2 = \bm{0}$, where $\bm{0}$ is the zero matrix. 
Therefore, the quantum spin-$1$ vdP oscillator ($J = 1$) is the smallest possible model of the general quantum spin vdP  {oscillator} defined by Eq.~\eqref{eq:qsvdp}. 

When the spin number is large ($J \gg 1$), the deterministic dynamics in the classical limit of Eq.~(\ref{eq:qsvdp}),
described by the drift term of the approximate semiclassical Fokker-Planck equation (FPE) derived from Eq.~(\ref{eq:qsvdp})
valid for $\gamma_2 \ll \gamma_1$, is given by
\begin{align}
	\label{eq:qsvdp_ctraj}
	\dot{u} = \left( \frac{\gamma_1}{2} - i \omega \right)u
	-  2J \gamma_2  u \abs{u}^2.
\end{align}
This is a normal form of a supercritical Hopf bifurcation (the Stuart-Landau oscillator), which can be derived from general oscillator models including the classical vdP  oscillator with a weak nonlinearity (see Appendix \ref{ap_semi}) and has a stable circular limit cycle on the complex plane~\cite{kuramoto1984chemical}.

We can show that, in the high-spin limit $J \to \infty$, this model becomes equivalent to the quantum optical vdP  {oscillator} \cite{lee2013quantum} (see Appendix \ref{ap_cres} for the derivation).
%

%%%%%%%%%%%%%%%%%
%%% Section 3 %%%
%%%%%%%%%%%%%%%%%
%
%\section{Results}
%
\begin{figure} [!t]
	\begin{center}
		\includegraphics[width=\hsize,clip]{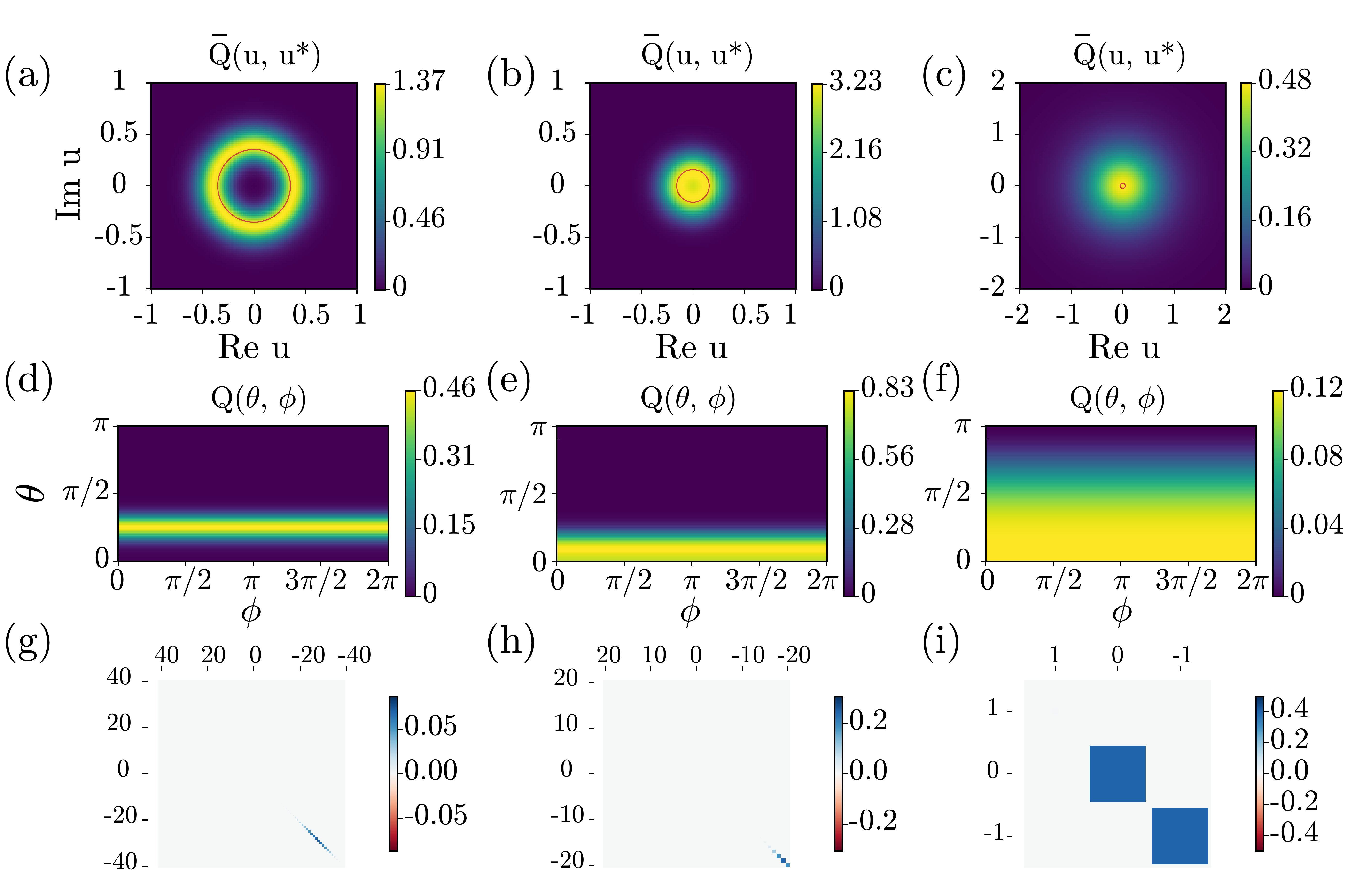}
		\caption{
			Limit-cycle oscillations of the quantum spin vdP oscillators with different spin numbers $J$. 
			(a-c): $\bar{Q}(u, u^*)$ functions on the complex plane 
			with the corresponding classical limit cycles of Eq.~(\ref{eq:qsvdp_ctraj}) (red-thin lines).
			(d-f): $Q(\theta, \phi)$ functions on the unit sphere.
			(g-i): Elements of the steady-state density matrices $\rho_{ss}$.
			The spin numbers and other parameters are
			(a, d, g) $J = 40$ and $(\omega, \gamma_2)/\gamma_1 = (-0.1, 0.05)$, 
			(b, e, h) $J = 20$ and $(\omega, \gamma_2)/\gamma_1 = (-0.1, 0.5)$,
			and (c, f, i) $J = 1$ and $(\omega, \gamma_2)/\gamma_1 =  (-0.1, 100)$,
			where $\gamma_1 = 1$. 
		}
		\label{fig_2}
	\end{center}
\end{figure}

\section{Stable limit-cycle oscillations in the phase space}
First, we illustrate the limit-cycle oscillations of the quantum spin vdP oscillator without external perturbations. Figure~\ref{fig_2} shows the steady state  $\rho_{ss}$ of Eq.~(\ref{eq:qsvdp})  for three different values of the spin number $J$ and other parameter sets; Figs.~\ref{fig_2}(a)-\ref{fig_2}(c) depict the functions $\bar{Q}(u, u^*)$ on the complex plane, Figs.~\ref{fig_2}(d)-\ref{fig_2}(f) depict the functions $Q(\theta, \phi)$ on the unit sphere,  and Figs.~\ref{fig_2}(g)-\ref{fig_2}(i) show the elements of the steady-state density matrices $\rho_{ss}$.
In Figs.~\ref{fig_2}(a)-\ref{fig_2}(c), we also show the stable limit-cycle solutions of Eq.~(\ref{eq:qsvdp_ctraj}) in the classical limit (see Appendix \ref{ap_semi}).

We can observe that the function $\bar{Q}(u, u^*)$ is localized around the classical limit cycle of Eq.~(\ref{eq:qsvdp_ctraj}) in all cases.
Note that the quantum limit-cycle oscillation is more clearly observed in Fig.~\ref{fig_2}(a) with a high spin number $J$ and a small nonlinear damping parameter $\gamma_2$, where the system is in the semiclassical regime. The limit-cycle oscillation is also apparent in the $(\theta, \phi)$-representation as shown in Figs.~\ref{fig_2}(d)-\ref{fig_2}(f), where $Q(\theta, \phi)$ is localized around a certain value of $\theta$,
while uniformly distributed without any preference for the phase $\phi$ \cite{roulet2018synchronizing}. 

The steady-state density matrices in Figs.~\ref{fig_2}(g)-\ref{fig_2}(i) take non-zero values only on the diagonal elements, reflecting that the system has 
no preference for specific phase values. Additionally, the diagonal elements of each density matrix are concentrated around the elements $\ket{J,m}\bra{J,m}$ with relatively small secondary spin quantum numbers $m$, and the element $\ket{J,J}\bra{J,J}$ with the largest secondary spin quantum number, $m = J$, takes tiny values close to zero. This indicates that the system rarely reaches the highest spin state (See Appendix \ref{ap_puri} for more details on the dependence on the spin number $J$).

\section{Frequency entrainment to an external drive}
Next, we consider the frequency entrainment of a single quantum spin vdP oscillator to a periodic external drive. 
We introduce the external drive by adding the following Hamiltonian to the original Hamiltonian $H$ in Eq.~(\ref{eq:qsvdp}):
\begin{align}
	\label{eq:ex}
	H_{ex}=i \frac{E}{2}\left(e^{i \omega_d t} J'_{-}-e^{-i \omega_d t} J'_{+} \right),
\end{align}
where $E$ and $\omega_d$ represent the intensity and frequency of the external drive, respectively \cite{roulet2018quantum}.
This external drive becomes equivalent to the harmonic drive for the quantum optical vdP oscillator 
in the high-spin limit \cite{arecchi1972atomic}.
In the rotating frame of the external drive frequency $\omega_d$,
the master equation for the oscillator under the harmonic drive can be derived as  \cite{roulet2018synchronizing}
\begin{align}
	\label{eq:qsvdp_ex}
	\dot{\rho}
	&= -i[-\Delta J_{z} + E J'_{y}, \rho] 
	+ \gamma_1 \mathcal{D}[J'_{+}]\rho
	+ \gamma_2 \mathcal{D}[(J'_{-})^2]\rho,
\end{align}
where $\Delta = \omega_{d} - \omega$ represents the frequency detuning of the external drive from the oscillator and the rescaled operator $J'_y = J_y/\sqrt{2J}$ was introduced.

To characterize the frequency entrainment, we use the observed frequency 
%$
defined as the peak frequency of the power spectrum in the steady state, 
\begin{align}
	\label{eq:obs}
	&
	\omega_{obs} = \argmax_{\omega} S(\omega), 
	\cr
	&
	S(\omega) = \int_{-\infty}^{\infty} d\tau e^{i\omega \tau} \left(\mean{ J'_{+}( \tau ) J'_{-}(0)} - \mean{J'_{+}( \tau )} \mean{J'_{-}(0)} \right),
\end{align}
where $\mean{A}= \Tr{[ A \rho_{ss}]}$ represents the expectation value of $A$ with respect to the steady state $\rho_{ss}$ of Eq.~\eqref{eq:qsvdp_ex} and $\mean{ J'_{+}( \tau ) J'_{-}(0)}$ is the correlation function of $J'_{+}(\tau)$ and $J'_{-}(0)$.
Here,  $\mean{ J'_{+}( \tau )}$ represents the expectation value of the operator $J'_{+}$ at time $\tau$ in the Heisenberg picture, 
where the expectation value is calculated using quantum regression theorem \cite{carmichael2007statistical}. In the high-spin limit, this power spectrum analysis becomes identical to that of quantum optical systems \cite{walter2014quantum, kato2019semiclassical}.
We also introduce the following {one-body} order parameter to evaluate the phase coherence of the oscillator:
%
%\begin{align}
%\label{eq:s1}
	$S_1  = \abs{S_1} e^{i\phi_1} 
	= \frac{ \langle J'_{-} \rangle }
	{\sqrt{  \langle J_{z} + J  \rangle } }$.
%\end{align}
%
The absolute value $\abs{S_1}$ represents the degree of the phase coherence and $\phi_1$ characterizes the mean phase of the system.
This quantity becomes equivalent to the order parameter $S^a_{1} = \frac{\langle a \rangle}{\sqrt{ \langle a^\dag a \rangle }}$  
used for the quantum optical systems \cite{lorch2016genuine, weiss2016noise} in the high-spin limit \footnote{If $\rho_{ss} = \ket{u}\bra{u}$, we obtain $S_1 = \frac{\sqrt{2J}u/(1+\abs{u}^2)}{\sqrt{2J\abs{u}^2/(1+\abs{u}^2)}} = \frac{u}{\sqrt{\abs{u}^2}\sqrt{1+\abs{u}^2}}$, and denoting $\alpha = re^{i\phi}$, it converge to $\frac{\alpha}{\sqrt{\abs{\alpha}^2}} = e^{i\phi}$ in the high-spin limit. }.

\begin{figure} [!t]
	\begin{center}
		\includegraphics[width=\hsize,clip]{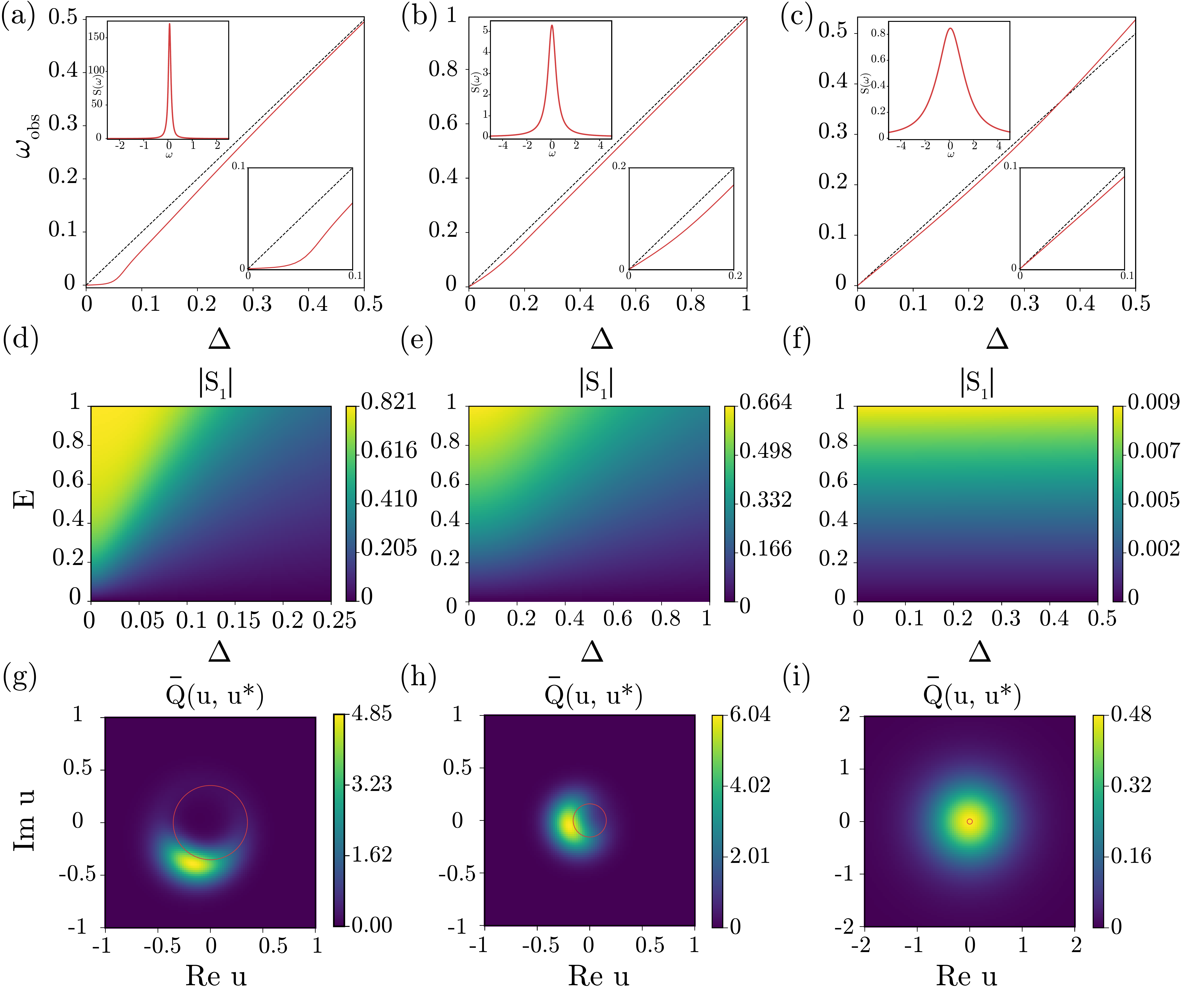}
		\caption{
			Frequency entrainment of a quantum spin vdP oscillator to a periodic external drive.
			(a-c): Observed frequencies $\omega_{obs}$. 
			The inset at top left in each figure displays the power spectrum when $\Delta = 0.1$.
			The inset at bottom right in each figure highlights the area with small detuning parameter $\Delta$. 
			(d-f): Order parameters $\abs{S_{1}}$ on the $\Delta-E$ plane, 
			(g-i): $\bar{Q}(u, u^*)$ functions  on the complex plane with stable limit-cycle solutions of Eq.~(\ref{eq:qsvdp_ctraj}) (red-thin lines).
			The parameters are 
			(a, d, g) $J = 40$ and $\gamma_2/\gamma_1 = 0.05$, 
			(b, e, h) $J = 20$ and $\gamma_2/\gamma_1 = 0.5$, 
			(c, f, i) $J = 1$ and $\gamma_2/\gamma_1 = 100$,
			(a, b, c) $E/\gamma_1 = 1$, 
			and
			(g, h, i) $(E, \Delta)/\gamma_1 = (1, 0.1)$,
			where $\gamma_1 = 1$.
		}
		\label{fig_3}
	\end{center}
\end{figure}

Figure~\ref{fig_3} shows the frequency entrainment of the quantum spin vdP oscillator for the same three parameter sets as in Figs.~\ref{fig_2}(a,d,g), \ref{fig_2}(b,e,h), and \ref{fig_2}(c,f,i); Figs.~\ref{fig_3}(a-c) plot the observed frequencies $\omega_{obs}$ with respect to the detuning parameter $\Delta$, Figs.~\ref{fig_3}(d-f) show the order parameters $\abs{S_{1}}$ on the $\Delta$-$E$ parameter plane, and Figs.~\ref{fig_3}(g-i) show the $\bar{Q}$-distributions $\bar{Q}(u, u^*)$ on the complex plane with the classical limit cycle of Eq.~(\ref{eq:qsvdp_ctraj}).

For the cases with large spin numbers, $J = 40$ in Figs.~\ref{fig_3}(a,d,g) and $J = 20$ in Figs.~\ref{fig_3}(b,e,h), we can clearly observe the frequency entrainment from the decrease in the observed frequency $\omega_{obs}$ in Figs.~\ref{fig_3}(a,b), Arnold tongues in Figs.~\ref{fig_3}(d,e), and the steady-state $\bar{Q}$ distributions localized around the phase-locking point in the classical limit in Figs.~\ref{fig_3}(g,h). The case with the larger spin number $J=40$ shows stronger tendency toward frequency entrainment than the case with $J=20$, as can be observed from the sharper decrease in the frequency difference between the oscillator and the external drive in Fig.~\ref{fig_3}(a) and narrower Arnold tongue in Fig.~\ref{fig_3}(d).

In contrast, the frequency entrainment is much weaker for the smallest spin-$1$ vdP oscillator ($J = 1$); 
we cannot observe the Arnold tongue in Fig.~\ref{fig_3}(f) nor phase-locked $\bar{Q}$ distribution in Fig.~\ref{fig_3}(i) clearly 
due to the strong quantum noise. We can only see a small tendency to frequency entrainment from the slight decrease in 
the observed frequency $\omega_{obs}$ for small detuning parameter $\Delta$ in Fig.~\ref{fig_3}(c).
However, $\omega_{obs}$ increases for larger $\Delta$, which never occurs in the quantum optical vdP oscillator.
This is because, unlike the quantum optical vdP oscillator, the quantum spin vdP oscillator can reach the highest spin state by the energy pumping from the external drive, which has a particularly strong effect on the dynamics of the smallest spin-$1$ vdP oscillator.  
In Appendix \ref{ap_deep}, we also present a perturbative analysis of the weak external drive that highlights the differences in the synchronization properties between in the quantum spin vdP system and quantum optical vdP system in the deep quantum limit.

\begin{figure} [!t]
	\begin{center}
		\includegraphics[width=1\hsize,clip]{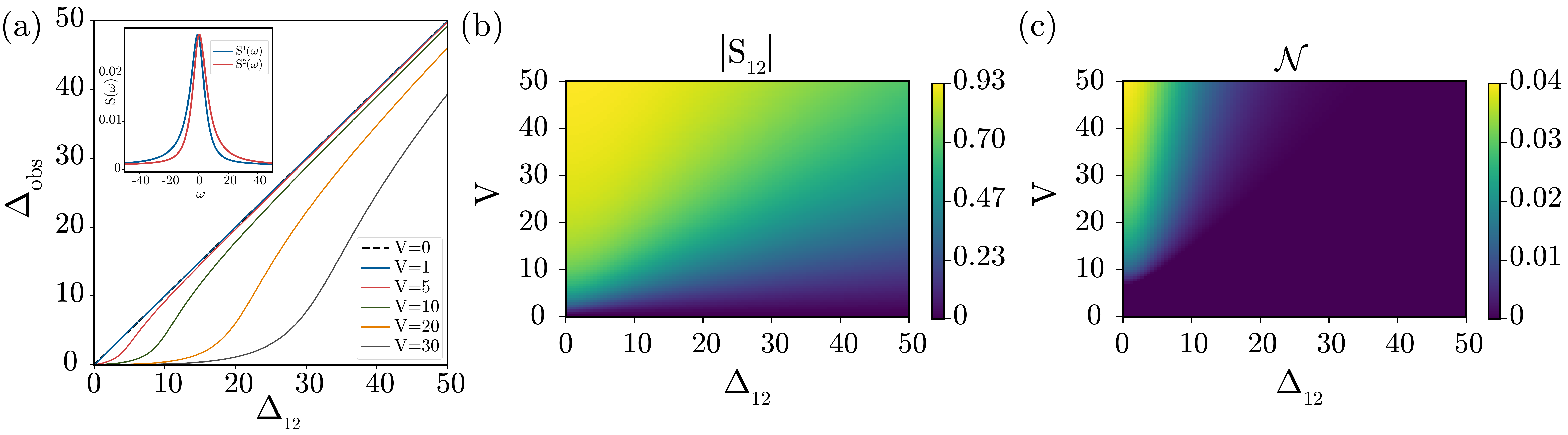}
		\caption{
			Frequency synchronization of two dissipatively coupled quantum spin vdP oscillators. 
			The inset displays the power spectra with $\Delta_{12} = 20, V= 30$.
			(a): Dependence of the observed detuning $\Delta_{obs}$  on the parameter $\Delta_{12}$.
			(b, c): Dependence of (b) the order parameter $\abs{S_{12}}$ and (c) the negativity $\mathcal{N}$
			on the parameters $V$ and $\Delta_{12}$.
			The other parameters are $J = 1$ and $\gamma_2/\gamma_1 = 100$.
			We set $\gamma_1 = 1$.
		}
		\label{fig_4}
	\end{center}
\end{figure}

\section{Frequency synchronization of two dissipatively coupled oscillators}
Next, we consider a pair of dissipatively coupled quantum spin vdP oscillators. 
The quantum master equation for the density matrix $\rho$ of this coupled system is {generally} given by
\begin{align}
	\label{eq:qsvdp2}
	\dot{\rho}
	&= 
	\sum_{k=1,2} 
	\left(
	-i[\omega_{k} J^{k}_{z}, \rho] 
	+ \gamma_{1} \mathcal{D}[J'^{k}_{+}]\rho
	+ \gamma_{2} \mathcal{D}[(J'^{k}_{-})^2]\rho
	\right)
	\cr
	&+ V \mathcal{D}[J'^{1}_{-} -  J'^{2}_{-}],
\end{align}
where $J^{k}_{z}$, $J'^{k}_{+}$, and $J'^{k}_{-}$ represent the (rescaled) angular momentum operators $J_z, J'_{+}$ and $J'_{-}$ for the $k$-th oscillator~$(k=1,2)$, $\omega_k$ represents the natural frequency of the $k$-th oscillator, and $V$ represents the strength of the dissipative coupling. We can show that, in the high-spin limit, this system becomes equivalent to a pair of dissipatively coupled quantum optical vdP oscillators \cite{lee2014entanglement, walter2015quantum}.

We define the observed frequency of each oscillator as the peak frequency of the power spectrum, 
\begin{align}
	\label{eq:obs2}
	\omega^{k}_{obs} 
	= \argmax_{\omega} S^{k}(\omega), 
	%\quad
	S^{k}(\omega) = \int_{-\infty}^{\infty} d\tau e^{i\omega \tau} 
	\left(\mean{ J'^{k}_{+}( \tau ) J'^{k}_{-}(0)} - \mean{J'^{k}_{+}( \tau )} \mean{J'^{k}_{-}(0)} \right),
\end{align}
and use the observed detuning $\Delta_{obs} = \omega^{2}_{obs} - \omega^{1}_{obs}$
to evaluate the frequency synchronization of the two oscillators.
We also introduce a {two-body} order parameter for evaluating the phase coherence between the two oscillators,
\begin{align}
	\label{eq:s12}
	S_{12} = \abs{S_{12}}e^{i \phi_{12}} = 
	\frac{\langle J'^{1}_{+}  J'^{2}_{-}  \rangle}
	{ \sqrt{  \langle J^{1}_{z} + J  \rangle \langle J^{2}_{z} + J \rangle }},
\end{align}
which is, in the high-spin limit,  equivalent to the quantum analog of the order parameter 
$S^a_{12} = 
\frac{\langle a_1^{\dag} a_2 \rangle}{\sqrt{ \langle a_1^\dag a_1 \rangle }\sqrt{ \langle a_2^\dag a_2 \rangle }}$  
for quantum optical systems~\cite{weiss2016noise}.
Here, the absolute value $\abs{S_{12}}$ represents the intensity of the phase coherence 
and $\phi_{12}$ characterizes the mean phase difference between the two oscillators.

To quantify the entanglement between the two oscillators, we use the negativity
\begin{align}
	\label{eq:negativity}
	\mathcal{N} = \frac{\left\|\rho^{\Gamma_{1}}\right\|_{1}-1}{2},
\end{align}
where $\rho^{\Gamma_{1}}$ represents the partial transpose of $\rho$ with respect to the subsystem of the oscillator $1$ and $\left\|X\right\|_{1}=\operatorname{Tr}|X|=\operatorname{Tr} \sqrt{X^{\dagger} X}$. The two oscillators are entangled 
when the negativity takes nonzero values \cite{zyczkowski1998volume, vidal2002computable}. 

Figure~\ref{fig_4} shows the results for the two dissipatively coupled spin-$1$ vdP oscillators 
with the smallest possible spin number; Fig.~\ref{fig_4}(a) shows the dependence of $\Delta_{obs}$ on the frequency detuning 
$\Delta_{12} {= \omega_2 - \omega_1}$, and Figs.~\ref{fig_4}(b) and \ref{fig_4}(c) show the order parameter $\abs{S_{12}}$ and negativity $\mathcal{N}$, respectively, with respect to the detuning $\Delta_{12} $ and coupling strength $V$. Here, we set $\omega_1 = \Delta_{12}/2$ and $\omega_2 = -\Delta_{12}/2$ in the numerical simulations. In Fig.~\ref{fig_4}(a), the frequency synchronization can be observed from the decrease in the observed detuning $\Delta_{obs}$, showing a stronger tendency toward frequency synchronization when the dissipative coupling strength $V$ is larger.

In {remarkable} contrast to the previous case for a single oscillator, we can clearly observe frequency synchronization, the Arnold tongue, and also the entanglement tongue~\cite{lee2014entanglement} between the two oscillators even for the smallest spin-$1$ case, similar to those observed for two dissipatively coupled quantum optical vdP oscillators~\cite{walter2015quantum, lee2014entanglement}. This is because the dissipative coupling tends to bring the two systems to lower energy levels and prevents both systems from reaching the highest spin states, making the system dynamics unaffected by the finite dimensionality even in the case of the smallest spin-$1$ vdP oscillator. We also note that the above result is the first explicit observation of mutual frequency synchronization between a pair of {\it single} quantum {spin-based} limit-cycle oscillators, although macroscopic effects of quantum frequency synchronization between two networks of globally coupled quantum limit-cycle oscillators have been discussed in \cite{nadolny2023macroscopic}. 

\section{Collective synchronization transition in a network of globally coupled oscillators}
Finally, we analyze collective synchronization in a population of globally coupled quantum spin-$1$ vdP oscillators.
We employ a mean-field approach similar to \cite{nadolny2023macroscopic} for globally coupled quantum spin-based oscillators.

We consider a large network of globally coupled $N$ quantum spin-$1$ vdP oscillators with identical properties. As in \cite{lee2014entanglement} for the quantum optical system, we introduce an ansatz that the density matrix of the whole system can be described as a product of the subsystems as $\rho=\bigotimes_n \rho_n$ in the $N \to \infty$ limit, where $\rho_n$ represents the density matrix of the $n$-th oscillator. Because all oscillators are identical, the density matrix of the coupled system obeys an equation with a mean-field coupling (see Appendix \ref{ap_mean} for the derivation), 
\begin{align}
	\label{eq:qsvdpn_sc}
	\dot{\rho}
	&= -i[\omega J_{z}, \rho] 
	+ \gamma_{1} \mathcal{D}[J'_{+}]\rho
	+ \gamma_{2} \mathcal{D}[(J'_{-})^2]\rho
	\cr
	& + K  \mathcal{D}[J'_{-}]\rho
	+ \frac{K}{2} \left( A^*[\rho, J'_{-}]+A[J'_{+}, \rho]\right),
\end{align}
where $\omega$ determines the frequency and $K$ denotes the strength of the global coupling, and $A = \Tr [J'_{-} \rho]$ and $A^{*} = \Tr [J'_{+} \rho]$ represent the mean fields of $J'_{+}$ and $J'_{-}$, respectively.

\begin{figure}[!t]
	\begin{center}
		\includegraphics[width=0.55\hsize,clip]{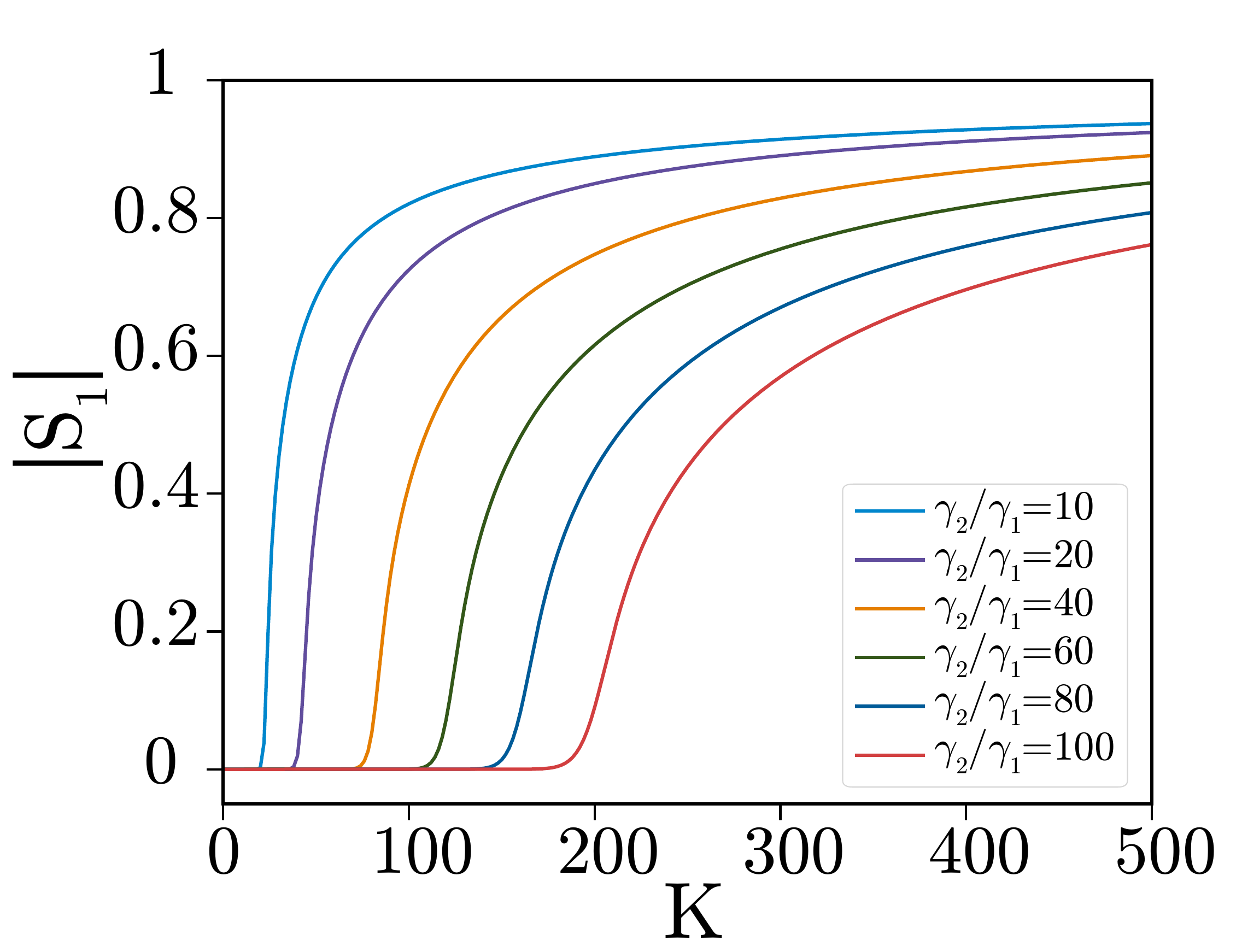}
		\caption{
			Collective synchronization of globally coupled quantum spin vdP oscillators.
			Dependence of the phase coherence $\abs{S_1}$ on the coupling strength $K$.   
			The parameters are $J = 1$ and $\omega= -20$.
			We set  $\gamma_1 = 1$. 
		}
		\label{fig_5}
	\end{center}
\end{figure}

Figure \ref{fig_5} illustrates the dependence of the phase coherence $\abs{S_1}$ on the coupling strength $K$ for several values of the nonlinear damping parameter $\gamma_2$. These values of $\gamma_2$ are sufficiently large that the finite dimensionality of the smallest spin-$1$ vdP oscillator does not significantly affect the system dynamics. We can clearly observe collective synchronization transitions from incoherent to coherent dynamics, i.e., sudden increase in $|S_1|$, around $K =  20, 40, 75, 110, 145$, and $180$ for $\gamma_2/\gamma_1 = 10, 20, 40, 60, 80$, and $100$, respectively. The critical coupling strength $K$ at which the synchronization transition occurs increases with 
$\gamma_2$. Thus, the quantum analogue of the Kuramoto transition \cite{kuramoto1984chemical} can also be observed clearly in globally coupled quantum spin-$1$ vdP oscillators. 

\section{Possibility of experimental implementation}
We consider that the quantum spin vdP oscillator proposed in this study can be experimentally implemented with the currently available physical setups. Experimental realization of phase synchronization in spin-$1$ systems was recently reported in an ensemble of Rb
atoms \cite{laskar2020observation} and on the IBM Q system \cite{koppenhofer2020quantum}. Similarly, quantum synchronization of spin-$J$ systems would also be achievable in collective ensembles of qubits using such systems. The assumed dissipative negative damping and nonlinear damping could be experimentally engineered  using the multi-level structures of atomic ensemble systems \cite{poyatos1996quantum, lutkenhaus1998mimicking}, as originally discussed in \cite{roulet2018synchronizing}.
The ion trap system, considered as a candidate system for the potential realization of the quantum optical vdP oscillator~\cite{lee2013quantum}, could also be used for experimental realization of the present quantum spin vdP oscillator.

\section{Conclusion}
We proposed the quantum spin vdP oscillator as a prototypical model of limit-cycle oscillators in quantum spin systems.
We demonstrated that it exhibits
all representative features, i.e.,
stable limit-cycle oscillations, frequency entrainment, mutual synchronization, and collective synchronization, 
which are essentially important in the analysis of classical synchronization.  We believe that the quantum spin vdP oscillator will provide valuable insights into novel features of quantum synchronization in many-body spin systems and future applications of quantum synchronization in the advancing fields of quantum technologies.\\

{\it Acknowledgments.-}
Numerical simulations are performed by using QuTiP numerical
toolbox \cite{johansson2012qutip, johansson2013qutip}. 
We acknowledge JSPS KAKENHI 
JP22K14274, JP22K11919, JP22H00516, 
25H01468, 25K03081, and
JST PRESTO JP-MJPR24K3
for financial support. 

\appendix

\clearpage

\newpage

%\setcounter{figure}{0}
%\renewcommand{\thefigure}{S\arabic{figure}}
%%%%%%%

%\setcounter{equation}{0}
%\renewcommand{\theequation}{S\arabic{equation}}
%%%%%%%

%
\section{Semiclassical approximation and the classical limit}
\label{ap_semi}
In this appendix, we derive Eq.~(\ref{eq:qsvdp_ctraj}) in the main text, which describes the deterministic dynamics in the classical limit of the quantum master equation~(\ref{eq:qsvdp}). Defining $p = m + J$ and $\ket{p} = \ket{J, p-J}$ in Eq.~\eqref{eq:scs_u_sm}, we obtain
\begin{align}
	&\ket{u} = 
	\sum_{p=0}^{2J} \binom{2J}{p}^{1/2} 
	\frac{u^{p}}{(1 + \abs{u}^2)^J} \ket{p},
\end{align}
where
\begin{align}
	&J_{+}\left| p \right\rangle = 
	\sqrt{\left( 2J - p \right)\left( p + 1 \right)}\left|p+1 \right\rangle,
	\cr
	&J_{-}\left| p \right\rangle = 
	\sqrt{\left( 2J - p + 1 \right) p}\left|p - 1 \right\rangle,
	\cr
	&J_{z}| p \rangle =(p-J)| p \rangle.
\end{align}
We define a projector $\Lambda$ to the spin coherent state and an weighted operator $\Lambda'$ as
\begin{align}
	\Lambda = \ket{u}\bra{u} = \sum_{p,q=0}^{2J} \Gamma_{p,q},
	\quad
	\Lambda' = \sum_{p,q=0}^{2J} p\Gamma_{p,q},
\end{align}
where 
\begin{align}
	\Gamma_{p,q} = %\sum_{p, q=0}^{2J} 
	\binom{2J}{p}^{1/2}  \binom{2J}{q}^{1/2} 
	\frac{u^{p} (u^{*})^q }{(1 + \abs{u}^2)^{2J}} \ket{p} \bra{q}.
\end{align}

With straightforward calculations, we obtain 
\begin{align}
	J^+ \Lambda = u^{-1} \Lambda',
	\quad 
	J^- \Lambda = u\left( 2J\Lambda - \Lambda'\right),
	\quad
	J^z \Lambda = \Lambda' - J \Lambda,
	\quad
	\frac{\partial}{\partial u} \Lambda = \frac{-2u^*J}{1+\abs{u}^2}  \Lambda 
	+ u^{-1}  \Lambda',
\end{align}
which lead to 
\begin{align} 
	\label{eq:cspd_ope}
	J_{+} \ket{u} \bra{u}
	&= \left[ \frac { \partial } { \partial u }  + \frac{2u^* J}{1 + \abs{u}^2} \right] \ket{u} \bra{u},
	\cr
	J_{-} \ket{u} \bra{u}
	&= \left[ - u^2 \frac { \partial } { \partial u }  + \frac{2 u J}{1 + \abs{u}^2} \right] \ket{u} \bra{u},
	\cr
	J_{z} \ket{u} \bra{u}
	&= \left[ u \frac { \partial } { \partial u }  + \frac{ J( \abs{u}^2 - 1)}{1 + \abs{u}^2} \right] \ket{u} \bra{u}.
\end{align} 

We introduce the $P$ quasiprobability functions \cite{arecchi1972atomic}
on the unit sphere and on the complex plane as 
\begin{align}
	\rho = \int d\theta d\phi  \sin \theta P ( \theta,\phi) \ket{\theta, \phi} \bra{\theta, \phi}
	= \int d^2 u \bar{P}(u,u^*)\ket{u}\bra{u},
	\quad
	\bar{P}(u, u^*) = 4 \frac{P(u,u^*)}{(1 +  \abs{u}^2)^2}, 
\end{align}
which satisfy
\begin{align}
	\int d\theta d\phi \sin \theta P ( \theta , \phi)
	= \int d^2 u \bar{P}(u,u^*) = 1.
\end{align}
Here, we introduced the quasiprobability distribution function $\bar{P}(u, u^{*})$, which includes the coefficient arising from the coordinate transformation,  in addition to $P(u, u^*)$.

From Eq.~(\ref{eq:cspd_ope}), we obtain the following correspondence between the actions of the angular-momentum operators on the density matrix $\rho$ and the actions of the differential operators on the quasiprobability distribution $\bar{P}(u, u^*)$:
\begin{align} 
	\label{eq:cspd_p}
	J_{+} \rho
	&\leftrightarrow \left[ -\frac { \partial } { \partial u }  + \frac{2u^* J}{1 + \abs{u}^2} \right]
	\bar{P}(u,u^*),
	\cr
	J_{-} \rho
	&\leftrightarrow \left[ \frac { \partial } { \partial u }u^2   + \frac{2 u J}{1 + \abs{u}^2} \right]
	\bar{P}(u,u^*),
	\cr
	J_{z} \rho
	&\leftrightarrow \left[ - \frac { \partial } { \partial u }u  + \frac{ J( \abs{u}^2 - 1)}{1 + \abs{u}^2} \right]\bar{P}(u,u^*).
\end{align} 
The above correspondence, expressing the actions of angular-momentum operators in the $P$ representation, was obtained in~\cite{narducci1975phase} 
for the function $P(\theta, \phi)$ in the  $(\theta, \phi)$-coordinate representation.

Using Eq.~(\ref{eq:cspd_p}), the time evolution of the quasiprobability distribution $\bar{P}(u,u^*)$ is obtained in the following form:
\begin{align}
	\label{eq:qfpe}
	\frac{\pa \bar{P}(u,u^*, t)}{\pa t} &= 
	\Big[ 
	- \sum_{j=1}^{2} \partial^u_{j}  A_{j}(u,u^*)
	+ 
	\frac{1}{2} \sum_{j=1}^2 \sum_{k=1}^2 \partial^u_{j}\partial^u_{k} D_{jk}(u,u^*)
	+
	\cdots 
	\Big]\bar{P}(u,u^*, t),
\end{align}
where $A_j(u, u^*)$ represents the $j$th component ($j=1,2$) of a complex vector $\bm{A}(u, u^*) = (A_1(u,u^*), A_2(u,u^*)) \in \mathbb{C}^{2 \times 1}$ with $A_2(u, u^*) = A^*_{1}(u, u^*)$, $D_{jk}(u, u^*)$ represents the $(j, k)$-th component ($j,k=1,2$) of the symmetric diffusion matrix $\bm{D}(u,u^*) \in \mathbb{C}^{2 \times 2}$ with $D_{12}(u, u^*) = D_{21}(u, u^*)$, $\partial^u_1 = \partial/\partial u $ and $\partial^u_2 = \partial/\partial u^* $ are the complex partial derivatives with respect to the first and second arguments, i.e., $u$ and $u^*$, respectively, 
and $\left[ \cdots \right]$ represents the complex partial derivative terms higher than the second order.

By explicitly calculating each term in Eq.~\eqref{eq:qfpe}, the time evolution of $\bar{P}(u,u^*, t)$ for the quantum spin vdP oscillator  Eq.~(\ref{eq:qsvdp}) in the main text is obtained as 
\begin{align}
	\label{eq:qsvdp_fpe}
	\frac{\pa \bar{P}(u,u^*, t)}{\pa t} &= \left[ - \sum_{j=1}^{2} \partial^u_{j}  A_{j}(u,u^*)
	+ 
	\frac{1}{2} \sum_{j=1}^2 \sum_{k=1}^2 \partial^u_{j}\partial^u_{k} D_{jk}(u,u^*)
	\right.
	\cr
	&
	-\frac{\gamma_2}{(2J)^2} \left(\frac{\pa^3}{\pa u^2 \pa u^{*}} 
	\frac{2 u\abs{u}^6}{1 + \abs{u}^2}
	( \abs{u}^2 - 2J + 1)
	\right.
	\cr 
	& 
	\left.
	%\frac{\pa^3}{\pa^3 u}
	-\frac{\pa^3}{\pa u^3}
	\frac{ u^3}{1 + \abs{u}^2}
	( (2J+1)\abs{u}^2 - 2J + 1)
	+ H.c. \right)
	\cr
	&
	\left.
	+
	\frac{\gamma_2}{(2J)^2} \frac{1}{2}\left( - \frac{\pa^4}{\pa u^4}u^4 + \frac{\pa^4}{\pa u^2 \pa u^{*2}}\abs{u}^8+ H.c. \right)
	\right]\bar{P}(u,u^*, t), 
\end{align}
%
%with
where
\begin{align}
	\bm{A}(u,u^*)
	&= \left( \begin{matrix}
		\left (\frac{\gamma_1}{2J}  (J+1)  - i \omega \right)u
		-\frac{\gamma_2}{(2J)^2}\frac{1}{ (1 + \abs{u}^2)^3 } 2J(2J-1)  u \abs{u}^2 ( \abs{u}^4 + (2J+1)\abs{u}^2 + 2J)
		\\
		\left (\frac{\gamma_1}{2J}  (J+1)  + i \omega \right)u^* 
		-\frac{\gamma_2}{(2J)^2}\frac{1}{ (1 + \abs{u}^2)^3 } 2J(2J-1)  u^* \abs{u}^2 ( \abs{u}^4 + (2J+1)\abs{u}^2 + 2J)
		\\
	\end{matrix} \right)
\end{align}
and 
\small
\begin{align}
	\bm{D}(u,u^*) 
	= 
	\scalebox{0.8}{$\displaystyle
		\left( \begin{matrix}
			\frac{\gamma_1}{2J} u^2
			+\frac{\gamma_2}{(2J)^2}\frac{1}{ (1 + \abs{u}^2)^2 } 2J(2J-1)u^2( \abs{u}^4 + 4\abs{u}^2 - 1) 
			&
			\frac{\gamma_1}{2J}
			+\frac{\gamma_2}{(2J)^2}\frac{4\abs{u}^6}{ (1 + \abs{u}^2)^2 }  ( \abs{u}^4 + 2(-J+1)\abs{u}^2 + (-2J+1)^2 )
			\\
			\frac{\gamma_1}{2J}
			+\frac{\gamma_2}{(2J)^2}\frac{4\abs{u}^6}{ (1 + \abs{u}^2)^2 }  ( \abs{u}^4 + 2(-J+1)\abs{u}^2 + (-2J+1)^2 )
			& 
			\frac{\gamma_1}{2J} u^{*2}
			+\frac{\gamma_2}{(2J)^2}\frac{1}{ (1 + \abs{u}^2)^2 } 2J(2J-1)
			u^{*2}( \abs{u}^4 + 4\abs{u}^2 - 1) 
			\\
		\end{matrix} \right) $}.
\end{align}
\normalsize

Rescaling the complex variable as $u = \frac{u'}{\sqrt{2J}}$, we obtain the evolution equation for the quasiprobability distribution 
in the $\bar{P}$ representation of the rescaled complex variables $u'$ and $u'^*$  as 
\begin{align}
	\label{eq:qsvdp_fpe_rs}
	\frac{\pa \bar{P}(u',u'^*, t)}{\pa t} &= \left[ - \sum_{j=1}^{2} \partial^{u'}_{j}  A_{j}(u',u'^*)
	+ 
	\frac{1}{2} \sum_{j=1}^2 \sum_{k=1}^2 \partial^{u'}_{j}\partial^{u'}_{k} D_{jk}(u',u'^*)
	\right.
	\cr
	&
	-\frac{\gamma_2}{(2J)^2} \left(\frac{\pa^3}{\pa u'^2 \pa u'^{*}} 
	\frac{2 u' \frac{\abs{u'}^6}{(2J)^2}}{(1 + \frac{\abs{u'}^2}{2J})}
	( \frac{\abs{u'}^2}{2J} - 2J + 1)
	\right.
	\cr 
	& 
	-
	\left.
	\frac{\pa^3}{\pa u'^3}
	\frac{ u'^3}{1 + \frac{\abs{u'}^2}{2J}}
	( (2J+1)\frac{\abs{u'}^2}{2J} - 2J + 1)
	+ H.c. \right)
	\cr
	&
	\left.
	+
	\frac{\gamma_2}{(2J)^2} \frac{1}{2} \left(-\frac{\pa^4}{\pa u'^4}u'^4 + \frac{\pa^4}{\pa u'^2 \pa u'^{*2}}\frac{\abs{u'}^8}{(2J)^2}+ H.c. \right)
	\right]\bar{P}(u',u'^*, t)
\end{align}
with
\begin{align}
	\bm{A}(u',u'^*) &=
	\left( \begin{matrix}
		\left (\frac{\gamma_1}{2J}  (J+1)  - i \omega \right)u'
		-\frac{\gamma_2}{(2J)^2}\frac{1}{ (1 + \frac{\abs{u'}^2}{2J})^3 } (2J-1)  u' \abs{u'}^2 (\frac{\abs{u'}^4}{(2J)^2} + (2J+1)\frac{\abs{u'}^2}{2J} + 2J)
		\\
		\left (\frac{\gamma_1}{2J}  (J+1)  + i \omega \right)u'^* 
		-\frac{\gamma_2}{(2J)^2}\frac{1}{ (1 + \frac{\abs{u'}^2}{2J})^3 } (2J-1)  u'^* \abs{u'}^2 (\frac{\abs{u'}^4}{(2J)^2} + (2J+1)\frac{\abs{u'}^2}{2J} + 2J)
		\\
	\end{matrix} \right)
\end{align}
and 
\small
\begin{align}
	\bm{D}(u', u'^*) 
	= 
	\scalebox{0.8}{$\displaystyle
		\left( \begin{matrix}
			\frac{\gamma_1}{2J} u'^2
			+\frac{\gamma_2}{(2J)^2}\frac{1}{ (1 + \frac{\abs{u'}^2}{2J})^2 } 2J(2J-1)u'^2( \frac{\abs{u'}^4}{(2J)^2} + 4 \frac{\abs{u'}^2}{2J} - 1) 
			&
			\gamma_1
			+\frac{\gamma_2}{(2J)^2}\frac{4 \frac{\abs{u'}^6}{(2J)^2} }{(1 + \frac{\abs{u'}^2}{2J})^2 }  ( \frac{\abs{u'}^4}{(2J)^2} + 2(-J+1)\frac{\abs{u'}^2}{2J} + (-2J+1)^2 )
			\\
			\gamma_1
			+\frac{\gamma_2}{(2J)^2}\frac{4 \frac{\abs{u'}^6}{(2J)^2} }{ (1 + \frac{\abs{u'}^2}{2J})^2 }  ( \frac{\abs{u'}^4}{(2J)^2} + 2(-J+1)\frac{\abs{u'}^2}{2J} + (-2J+1)^2 )
			& 
			\frac{\gamma_1}{2J} u'^{*2}
			+\frac{\gamma_2}{(2J)^2}\frac{1}{ (1 + \frac{\abs{u'}^2}{2J})^2 } 2J(2J-1)
			u'^{*2}( \frac{\abs{u'}^4}{(2J)^2} + 4\frac{\abs{u'}^2}{2J} - 1) 
			\\
		\end{matrix} \right) $},
\end{align}
\normalsize
where the complex partial derivatives with respect  to $u'$ and $u'^*$ are defined as 
$\partial^{u'}_1 = \partial/\partial u' $ and $\partial^{u'}_2 = \partial/\partial u'^* $, respectively.

Assuming that the spin number $J$ is sufficiently large and neglecting the terms of order $\mathcal{O}(1/J)$ in Eq.~\eqref{eq:qsvdp_fpe_rs}, we obtain the approximate semiclassical FPE representing the time evolution of the $\bar{P}$ function as
\begin{align}
	\label{eq:qsvdp_fpe_re_apx}
	\frac{\pa \bar{P}(u',u'^*, t)}{\pa t} &= \left[ - \sum_{j=1}^{2} \partial^{u'}_{j}  A_{j}(u',u'^*)
	+ 
	\frac{1}{2} \sum_{j=1}^2 \sum_{k=1}^2 \partial^{u'}_{j}\partial^{u'}_{k} D_{jk}(u',u'^*)
	\right]\bar{P}(u',u'^*, t)
\end{align}
with
\begin{align}
	\bm{A}(u',u'^*) &=
	\left( \begin{matrix}
		\left ( \frac{\gamma_1}{2} - i \omega \right)u'
		- \gamma_2  u' \abs{u'}^2 
		\\
		\left (\frac{\gamma_1}{2}  + i \omega \right)u'^* 
		- \gamma_2 u'^* \abs{u'}^2 
		\\
	\end{matrix} \right)
\end{align}
and 
\begin{align}
	\bm{D}(u', u'^*) 
	= 
	\scalebox{0.8}{$\displaystyle
		\left( \begin{matrix}
			- \gamma_2 u'^2
			&
			\gamma_1
			\\
			\gamma_1
			& 
			- \gamma_2 u'^{*2}
			\\
		\end{matrix} \right) $},
\end{align}
which has the same form as the semiclassical FPE for the quantum optical vdP oscillator obtained in \cite{kato2019semiclassical}.

Now, further assuming $\gamma_2 \ll \gamma_1$ and neglecting the effect of quantum noise, 
we obtain the deterministic dynamics in the classical limit as 
\begin{align}
	\label{eq:qsvdp_ctraj_re_apx}
	\dot{u'} = \left( \frac{\gamma_1}{2} - i \omega \right)u'
	- \gamma_2  u' \abs{u'}^2, 
\end{align}
which is equivalent to the deterministic dynamics in the classical limit of the quantum optical vdP oscillator in Eq.~(\ref{eq:qvdp_me}) 
described in the next appendix.
Using the original complex variable $u' = \sqrt{2J} u$, we obtain 
\begin{align}
	\label{eq:qsvdp_ctraj_apx}
	\dot{u} = \left( \frac{\gamma_1}{2} - i \omega \right)u
	-  2J \gamma_2  u \abs{u}^2,
\end{align}
which gives the deterministic dynamics, Eq.~(\ref{eq:qsvdp_ctraj}), in the main text.
We note that, although the limit-cycle solution in the classical limit is derived via the $\bar{P}$ function, which is similar to but slightly different from the $\bar{Q}$ function, the function $\bar{Q}(u, u^*)$ localizes near the stable limit-cycle solution 
in the classical limit as shown in the figures in the main text. 
Note also that the phase-space representation is singular at the point $\theta = \pi$ (i.e., the north pole where $\phi$ is undefined) on the sphere in Eq.~(\ref{eq:scs_sm}), and we excluded this point in the above derivation  \cite{arecchi1972atomic}. 

We also note that in Ref. \cite{dutta2025quantum},  a different semiclassical approach has been employed to introduce limit-cycle oscillations  in the classical limit of quantum spin systems.

\section{Correspondence to the quantum optical vdP oscillator in the high-spin limit}
\label{ap_cres}
In this appendix, we show that the quantum spin vdP oscillator becomes equivalent to the quantum optical vdP oscillator in the high-spin limit ($J \to \infty$). In this limit, through the process known as the group contraction \cite{inonu1953contraction, segal1951class}, the operators for the spin systems can be related to those for the quantum optical systems as \cite{arecchi1972atomic}
\begin{align} 
	\label{eq:cspd}
	u \leftrightarrow \alpha/\sqrt{2J},
	\quad
	J_{-} \leftrightarrow \sqrt{2J} a, 
	\quad
	J_{+} \leftrightarrow \sqrt{2J} a^\dag,
	\quad
	J_{z} \leftrightarrow  a^\dag a - J,
\end{align} 
where $a$ and $a^\dag$ represent the annihilation and creation operators for the harmonic oscillator, respectively. By this correspondence, the spin coherent state in the high-spin limit becomes equivalent to the quantum optical coherent state in the harmonic-oscillator systems \cite{carmichael2007statistical, gardiner1991quantum}. Note that, though not used in this paper, we can also consider the correspondence in the high-spin limit of the Holstein-Primakoff transformation \cite{holstein1940field}, in which the lowering operator corresponds to the creation operator.

Using Eq.~(\ref{eq:cspd}), the quantum master equation (\ref{eq:qsvdp}) in the main text can be transformed to
\begin{align}
	\label{eq:qvdp_me}
	\dot{\rho} = 
	-i \left[  \omega a^{\dag}a, \rho \right]
	+ \gamma_{1} \mathcal{D}[a^{\dag}]\rho + \gamma_{2}\mathcal{D}[a^{2}]\rho
\end{align}
in the high-spin limit, which is the quantum master equation of a quantum optical vdP oscillator \cite{lee2013quantum}.
The time evolution of $\bar{P}(u,u^*, t)$ in Eq.~(\ref{eq:qsvdp_fpe}) can also be transformed to the time evolution 
of the $P$ distribution $\bar{P}(\alpha, \alpha^*, t)$ in the quantum optical coherent-state representation, 
\begin{align}
	\label{eq:qvdp_fpe}
	\frac{\pa P(\alpha, \alpha^*, t)}{\pa t} = \Big[ - \sum_{j=1}^{2} \partial^\alpha_{j} \{ A_{j}(\alpha, \alpha^*), t) \}
	+ \frac{1}{2} \sum_{j=1}^2 \sum_{k=1}^2 \partial^\alpha_{j}\partial^\alpha_{k} \{ D_{jk}(\alpha, \alpha^*) \} \Big]P(\alpha, \alpha^*, t)
\end{align}
with
\begin{align}
	\bm{A}(\alpha, \alpha^*)
	= \left( \begin{matrix}
		\left(\frac{\gamma_1}{2} - i \omega \right) \alpha  
		- \gamma_{2}\alpha \abs{\alpha}^{2} 
		\\
		\left(\frac{\gamma_1}{2} + i \omega \right) \alpha^{*}   
		- \gamma_{2} \alpha^{*}  \abs{\alpha}^{2} 
		\\
	\end{matrix} \right)
\end{align}
and 
\begin{align}
	\bm{D}(\alpha, \alpha^*) = 
	\left( \begin{matrix}
		-\gamma_{2}\alpha^{2}     &  \gamma_1  \\
		\gamma_1 & - \gamma_{2}\alpha^{*2} \\
	\end{matrix} \right),
\end{align}
where the complex partial derivatives are defined as 
$\partial^\alpha_1 = \partial/\partial \alpha $ and 
$\partial^\alpha_2 = \partial/\partial \alpha^* $.
This FPE is equivalent to the FPE for a quantum optical vdP oscillator \cite{kato2019semiclassical}.

The deterministic dynamics in the classical limit, i.e., the dynamics described by the drift vector field in Eq.~(\ref{eq:qvdp_fpe}), is given by
\begin{align}
	\label{eq:qvdp}
	\dot{\alpha} = 
	\left(\frac{\gamma_1}{2} - i \omega \right) \alpha  
	- \gamma_{2}\alpha^{*} \alpha^{2},
\end{align}
which is equivalent to the deterministic equation obtained in the classical limit of the quantum optical vdP oscillator described by Eq.~(\ref{eq:qvdp_me})~\cite{lee2013quantum, walter2014quantum, lee2014entanglement}. This equation represents the normal form of the supercritical Hopf bifurcation, which can be derived from the classical vdP oscillator with a weak nonlinearity \cite{van1927vii} near the Hopf bifurcation point.

Finally, we note that the nonlinear oscillator in Eq.~(\ref{eq:qvdp}) is known as the Stuart-Landau oscillator~\cite{kuramoto1984chemical} in the literature, and the quantum optical vdP oscillator in Eq.~(\ref{eq:qvdp_me}) has also been referred to as the quantum Stuart-Landau oscillator recently~\cite{chia2020relaxation,mok2020synchronization,wachtler2020dissipative}. Therefore, the quantum spin vdP oscillator proposed in this study may also be called the quantum spin Stuart-Landau oscillator.

\section{Dependence of purity on the spin number}
\label{ap_puri}

\begin{figure}[!t]
	\begin{center}
		\includegraphics[width=0.95\hsize,clip]{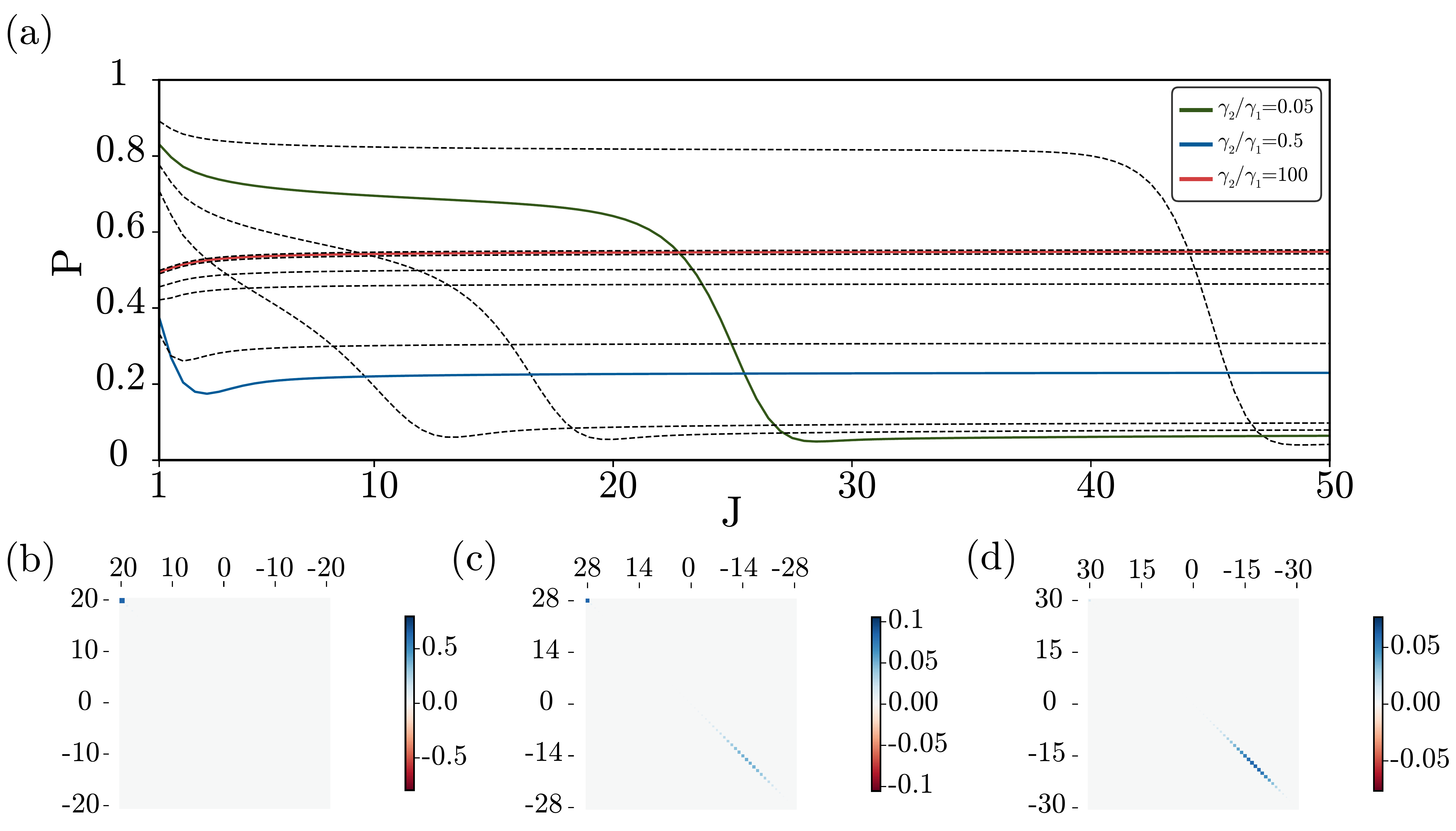}
		\caption{
			(a):
			Dependence of the purity $P$ on the spin number $J$.            
			$\gamma_2/\gamma_1 = 0.05$~(green-thin line), 
			$\gamma_2/\gamma_1 = 0.5$~(blue-thin line),
			$\gamma_2/\gamma_1 = 100$~(red-thin line), and 
			$\gamma_2/\gamma_1 = (0.03, 0.07, 0.1, 1, 5, 10, 50, 500)$~(black-dotted lines).
			(b-d):
			Elements of the density matrices $\rho_{ss}$ for 
			the parameters $(\omega, \gamma_2)/\gamma_1 = (-0.1, 0.05)$
			with (b) $J = 20$,  (c) $J = 28$, and (d) $J = 30$. 
			We set $\gamma_1 = 1$. 
		}
		\label{fig_6}
	\end{center}
\end{figure}

In this appendix, to investigate the dependence of the system dynamics on the spin number $J$, we examine the purity $P = \Tr[\rho_{ss}^2]$ in the steady state, which characterizes qualitative changes in the properties of the density matrix $\rho$. 

Figure~\ref{fig_6}(a) plots the purity $P$ as a function of the spin number $J$ for various values of the nonlinear damping parameter $\gamma_2$. When the nonlinear damping parameter $\gamma_2$ is relatively small, the purity $P$ undergoes a transition from a large value to a small value as $J$ increases.  For example, such transitions are observed in Fig.~\ref{fig_6}(a) around $J = J_0 = 28$ and $J_0= 2$ for the cases with $\gamma_2/\gamma_1 = 0.05$ and $0.5$, respectively. For a large nonlinear damping parameter $\gamma_2$, such as $\gamma_2/\gamma_1 = 100$, such a transition cannot be observed. We also illustrate density matrices near the transition points in Figs.~\ref{fig_6}(b-d) for $\gamma_2/\gamma_1 = 0.05$, where the values of $J$ are (b) $J = 20$, (c) $J=28$, and (d) $J =30$, respectively.

The reason for the transitions in the purity $P$ are as follows. Before the transitions, the effect of negative damping is relatively stronger than the effect of nonlinear damping, i.e. the energy pumping is relatively stronger than the energy dissipation, causing the diagonal elements of the density matrix to concentrate on the state $\ket{J,J}\bra{J,J}$. After the transitions, the effect of negative damping is relatively weaker than the effect of nonlinear damping,  i.e. the energy pumping is relatively weaker than the energy dissipation,  causing the diagonal elements of the density matrix to concentrate on the states $\ket{J,m}\bra{J,m}$ with relatively small spin numbers $m$. 

In Figs.~\ref{fig_2}(a)-(i) in the main text, we can observe the limit-cycle oscillation when the system's spin number $J$ is larger than the spin number $J_0$ at which the purity $P$ shows a transition in Fig.~\ref{fig_6}(a), whereas the limit-cycle oscillation cannot be observed clearly when the system's spin number $J$ is smaller than $J_0$. Thus, the spin number $J_0$ characterizes whether clear limit-cycle oscillations can be observed or not in the quantum spin vdP oscillator.

\section{Comparison to the quantum optical van der Pol oscillator in the deep quantum limit }
\label{ap_deep}

In the deep quantum limit $\gamma_2/\gamma_1 \to \infty$, the quantum optical van der Pol (vdP) oscillator can also be regarded as a three-level system~\cite{lee2013quantum, lee2014entanglement, mok2020synchronization}.
In this section, we consider this deep quantum limit and discuss the differences between the spin-$1$ vdP oscillator and the three-level quantum optical vdP oscillator.

The steady state of the quantum spin-$1$ vdP oscillator in Eq.~(\ref{eq:qsvdp}) in the main text with $J = 1$ is given by
\begin{align}
	\rho_{ss} = \displaystyle \frac{\gamma_{1}}{2\gamma_{2} + \gamma_{1}} \ket{1}\bra{1} + 
	\frac{\gamma_{2}}{2\gamma_{2} + \gamma_{1}} \ket{0}\bra{0} + 
	\frac{\gamma_{2}}{2\gamma_{2} + \gamma_{1}} \ket{-1}\bra{-1}.
\end{align}
In the deep quantum limit $\gamma_2/\gamma_1 \to \infty$, the steady state converges to a two-level system,
\begin{equation}
\label{eq:qsvdp_ss}
\rho_{ss} =  \displaystyle 
\frac{1}{2} \ket{0}\bra{0} + \frac{1}{2} \ket{-1}\bra{-1} .
\end{equation}

Let us also consider
the quantum spin-$1$ vdP oscillator 
subjected to a harmonic external drive
described by Eq.~(\ref{eq:qsvdp_ex}) in the main text, i.e., 
\begin{align}
	\label{eq:qsvdp_ex2}
	\dot{\rho}
	&= -i[-\Delta J_{z} + E J'_{y}, \rho] 
	+ \gamma_1 \mathcal{D}[J'_{+}]\rho
	+ \gamma_2 \mathcal{D}[(J'_{-})^2]\rho.
\end{align}
We assume that the 
intensity of the external drive $E$ is sufficiently small
and apply a perturbative approach~\cite{lee2013quantum, lee2014entanglement}.
Specifically, we consider a small 
variation $\delta \rho$
around the steady state $\rho_{ss}$,
i.e.,
$ \rho = \rho_{ss} + \delta \rho$,
where $ \delta \rho$ is assumed to be of the order $\mathcal{O} (E)$. 
We discuss the phase coherence, characterized by the off-diagonal elements of the density matrix. For simplicity, we assume that the frequency of the external drive matches the natural frequency of the system, i.e., $\Delta = 0$. In the following discussion, this assumption can be made without loss of generality since the detuning only induces a rotation of the symmetric system and does not affect phase coherence under a small perturbation.

Substituting $ \rho = \rho_{ss} + \delta \rho$ into Eq.~(\ref{eq:qsvdp_ex2}) and considering the steady state $\delta \rho_{ss}$ under small perturbations, i.e., assuming $\dot{\delta \rho}_{ss} = 0$ and neglecting small terms of order $\mathcal{O} (E^2)$, we can explicitly obtain the off-diagonal elements of the steady state  $ \delta \rho_{ss, 1, 0} (= \delta \rho_{ss, 0, 1})$, $\delta \rho_{ss, 0, -1} (= \delta \rho_{ss, -1, 0})$, $\delta \rho_{ss, 1, -1} (= \delta \rho_{ss, -1, 1}) $ as
\begin{align}
	\delta\rho_{1,0} &= - \left( \frac{E ( \gamma_2 - \gamma_1 )}{( \gamma_2 + \gamma_1 ) ( 2 \gamma_2 + \gamma_1 )} \right), \\
	\delta\rho_{ss,  0,-1} &= 0, \\
	\delta\rho_{ss,  1, -1} &= 0.
\end{align}
In the deep quantum limit, $\gamma_2/\gamma_1 \to \infty$, we obtain 
\begin{align}
	\delta\rho_{ss,  1. 0} \to 0,
	\quad
	\delta\rho_{ss,  0, -1} \to 0,
	\quad
	\delta\rho_{ss,  1, -1} \to 0,
\end{align}
which indicates that phase coherence $S_1$ 
vanishes at the first order perturbation
in this case. 

Next, we consider the quantum optical vdP systems. Given that we will take the deep quantum limit  $\gamma_2/\gamma_1 \to \infty$,  we restrict the system to the three levels $\ket{0}$, $\ket{1}$, and $\ket{2}$, and regard the annihilation operator as
$a =  \ket{0}\bra{1} + \sqrt{2} \ket{1}\bra{2}$. 
We note that, in this case, the energy levels $\ket{0}$, $\ket{1}$, and $\ket{2}$ correspond to the energy levels $\ket{-1}$, $\ket{0}$, and $\ket{1}$ for a spin-1 system.

The steady state of the quantum three-level optical vdP system is given by 
\begin{align}
	\rho_{ss} = \displaystyle \frac{\gamma_{1}}{3\gamma_{2} + \gamma_{1}} \ket{2}\bra{2} + 
	\frac{\gamma_{2}}{3\gamma_{2} + \gamma_{1}} \ket{1}\bra{1} + 
	\frac{2\gamma_{2}}{3\gamma_{2} + \gamma_{1}} \ket{0}\bra{0}.
\end{align}
	In the deep quantum limit $\gamma_2/\gamma_1 \to \infty$, the steady state converges to a two-level system
	\begin{equation}
		\label{eq:qvdp_ss}
		\rho_{ss} =  \displaystyle 
		\frac{1}{3} \ket{1}\bra{1} + \frac{2}{3} \ket{0}\bra{0}.
	\end{equation}
	Therefore, the steady state of the quantum spin van der Pol (vdP) system converges to the unweighted two-level system described by Eqs. (\ref{eq:qsvdp_ss}), whereas the quantum optical vdP system converges to the weighted two-level system in Eq.~(\ref{eq:qvdp_ss}). This discrepancy arises from the fundamental differences between the spin ladder operators and the optical annihilation and creation operators.

Let us also consider the quantum optical vdP system subjected to a harmonic external drive with frequency $\omega_d$.
	We consider the system corresponding to Eq.~(\ref{eq:qsvdp_ex}) in the main text, which, in the rotating frame of the external drive, is described by
	\begin{align}
		\label{eq:qvdp_ex}
		\dot{\rho} = 
		-i \left[  -\Delta a^{\dag}a - \frac{i E  }{2}(a^{\dag} - a), \rho \right]
		+ \gamma_{1} \mathcal{D}[a^{\dag}]\rho + \gamma_{2}\mathcal{D}[a^{2}]\rho,
	\end{align}
	where $\Delta = \omega_{d} - \omega$ represents the frequency detuning of the external drive from the oscillator.
	We assume that the 
	intensity 
	$E$
	of the harmonic drive  is sufficiently small
	and apply a perturbative approach~\cite{lee2013quantum, lee2014entanglement}.
	We consider a small 
	variation $\delta \rho$ 
	around the steady state $\rho_{ss}$,
	i.e., $ \rho = \rho_{ss} + \delta \rho$,
	where $ \delta \rho$ is assumed to be of the order $\mathcal{O} (E)$. 
	As in the spin vdP case,  we 
	focus on  the phase coherence characterized by the off-diagonal elements of the density matrix, and assume the frequency detuning $\Delta = 0$.
	
	Substituting $ \rho = \rho_{ss} + \delta \rho$ into Eq.~(\ref{eq:qvdp_ex}) and considering the steady state $\delta \rho_{ss}$ under small perturbations, i.e., assuming $\dot{\delta \rho}_{ss} = 0$ and neglecting small terms of of order $\mathcal{O} (E^2)$, we can explicitly obtain  the off-diagonal elements of the steady state,  $ \delta \rho_{ss, 2, 1} (= \delta \rho_{ss, 1, 2})$, $\delta \rho_{ss, 1, 0} (= \delta \rho_{ss, 0, 1})$, and $\delta \rho_{ss, 2, 0} (= \delta \rho_{ss, 0, 2})  $ as	
	\begin{align}
		\delta\rho_{ss,  2, 1} &= 
		-  \frac{ E ( 5 \gamma_2 - 3\gamma_1)}
		{3 \sqrt{2}(\gamma_2 + \gamma_1 ) ( 3 \gamma_2  + \gamma_1)},
		\\
		\delta\rho_{ss,  1, 0} &=  -  \frac{E\gamma_2}{3 \gamma_1 (3\gamma_2 + \gamma_1)},
		\\
		\delta\rho_{ss,  2, 0} &= 0.
	\end{align}
	In the high spin limit $\gamma_2/\gamma_1 \to \infty$, we obtain 
	\begin{align}
		\delta\rho_{ss,  2, 1} \to 0, 
		\quad
		\delta\rho_{ss,  1, 0} \to -  \frac{E}{9 \gamma_1} , 
		\quad
		\delta\rho_{ss,  2, 0} \to 0.
	\end{align}
	This indicates that the phase coherence $\abs{S_1}$ retains a non-zero value 
	%due to the small perturbation
	at the first order perturbation in this quantum optical vdP case,
	in contrast to the quantum spin vdP case.
	This difference in the phase coherence between the quantum spin system and the quantum optical system is caused by the difference in their respective stationary density matrices, especially the difference between weighted and unweighted mixed states. 

\section{Derivation of the mean-field equation for globally coupled oscillators}
\label{ap_mean}
In this appendix, we derive the mean-field equation (\ref{eq:qsvdpn_sc}) in the main text.  Our starting point is the following quantum master equation for the network of globally coupled $N$ quantum spin vdP oscillators: 

\begin{align}
	\label{eq:qsvdpN}
	\dot{\rho}
	&= 
	\sum_{k=1}^{N} 
	\left(
	-i[\omega J^{k}_{z}, \rho] 
	+ \gamma_{1} \mathcal{D}[J'^{k}_{+}]\rho
	+ \gamma_{2} \mathcal{D}[(J'^{k}_{-})^2]\rho
	\right)
	\cr
	& + \frac{K}{N} \sum_{k = 1}^{N} \sum_{k' = 1, k' < k}^{N} 
	\mathcal{D}[ (J'^{k'}_{-} -  J'^{k}_{-})]\rho,
\end{align}
where $J^{k}_{z}$, $J'^{k}_{+}$, and $J'^{k}_{-}$ represent the (rescaled) angular momentum operators for the $k$-th$~(k=1,2, \ldots, N)$ oscillator, $\omega_k$ represents the natural frequency of the $k$-th oscillator, and $K$ represents the strength of the dissipative coupling. Note that $\mathcal{D}[L] = \mathcal{D}[-L]$ for a given operator $L$.

We introduce the ansatz that the density matrix of the whole system is described as the product of the density matrices of the subsystems $\rho=\bigotimes_n \rho_n$ \cite{lee2014entanglement}. Then, by taking the partial trace over all $\rho_k'$ other than $\rho_k$ ($k' \neq k$),  
the time evolution of the $k$-th  subsystem $\rho_k$ is described as 
\begin{align}
	\label{eq:qsvdp_k}
	\dot{\rho}_k
	&= 
	\left(
	-i[\omega J^{k}_{z}, \rho_k] 
	+ \gamma_{1} \mathcal{D}[J'^{k}_{+}]\rho_k
	+ \gamma_{2} \mathcal{D}[(J'^{k}_{-})^2]\rho_k
	\right)
	\cr
	&+ \frac{K(N-1)}{N} \mathcal{D}[J'^{k}_{-}]\rho_k
	+ \frac{K}{2} \left( A_N^*[\rho_k, J'^{k}_{-}]+A_N[J'^{k}_{+}, \rho_k]\right),
\end{align}
where 
\begin{align}
	A_N = \frac{1}{N}\sum_{k'=1, k' \neq k}^{N}  \Tr [J'^{k}_{-} \rho_k],
	\quad
	A^*_N =  \frac{1}{N}\sum_{k'=1, k' \neq k}^{N}  \Tr [J'^{k}_{+} \rho_k],
\end{align}
are the averages of the expectation values of  $J'^{k}_{+}$ and $J'^{k}_{-}$  taken over all oscillators.

In the limit $N \to \infty$~\cite{spohn1980kinetic}, 
we obtain the mean-field equation in the form
\begin{align}
	\label{eq:qsvdp_mean}
	\dot{\rho}
	&= -i[\omega J_{z}, \rho] 
	+ \gamma_{1} \mathcal{D}[J'_{+}]\rho
	+ \gamma_{2} \mathcal{D}[(J'_{-})^2]\rho
	\cr
	& + K  \mathcal{D}[J'_{-}]\rho
	+ \frac{K}{2} \left( A^*[\rho, J'_{-}]+A[J'_{+}, \rho]\right),
\end{align}
where $A = \Tr [J'_{-} \rho]$ and $A^{*} = \Tr [J'_{+} \rho]$, and the oscillator index $k$ was dropped because all oscillators are statistically equivalent. 
Therefore, the globally coupled $N$ oscillators in Eq.~(\ref{eq:qsvdpN}) can be regarded as a single quantum system with a mean-field coupling representing the averaged interaction with the other systems, as in Eq.~(\ref{eq:qsvdp_mean}) or in Eq.~(\ref{eq:qsvdpn_sc}) in the main text.

%\bibliographystyle{apsrev4-1}
%\bibliographystyle{unsrt}
%\bibliography{reference_row}

\end{document}